# Sliding nanomechanical resonators

Yue Ying[1,2,†], Zhuo-Zhi Zhang[1,2,†], Joel Moser[3,4,*], Zi-Jia Su[1,2], Xiang-Xiang Song[1,2,*], and Guo-Ping Guo[1,2,5,*]

1. CAS Key Laboratory of Quantum Information, University of Science and Technology of China, Hefei, Anhui 230026, China
2. CAS Center for Excellence in Quantum Information and Quantum Physics, University of Science and Technology of China, Hefei, Anhui 230026, China
3. School of Optoelectronic Science and Engineering, Soochow University, Suzhou, Jiangsu 215006, China
4. Key Lab of Advanced Optical Manufacturing Technologies of Jiangsu Province, Soochow University, Suzhou, Jiangsu 215006, China
5. Origin Quantum Computing Company Limited, Hefei, Anhui 230088, China

† Y. Y. and Z.-Z. Z. contributed equally to this work.

* Author to whom correspondence should be addressed: J. M. (j.moser@suda.edu.cn), X.-X. S. (songxx90@ustc.edu.cn), or G.-P. G. (gpguo@ustc.edu.cn).

**Abstract**

The motion of a vibrating object is determined by the way it is held. This simple observation has long inspired string instrument makers to create new sounds by devising elegant string clamping mechanisms, whereby the distance between the clamping points is modulated as the string vibrates. At the nanoscale, the simplest way to emulate this principle would be to controllably make nanoresonators slide across their clamping points, which would effectively modulate their vibrating length. Here, we report measurements of flexural vibrations in nanomechanical resonators that reveal such a sliding motion. Surprisingly, the resonant frequency of vibrations draws a loop as a tuning gate voltage is cycled. This behavior indicates that sliding is accompanied by a delayed frequency response of the resonators, making their dynamics richer than that of resonators with fixed clamping points. Our work elucidates the dynamics of nanomechanical resonators with unconventional boundary conditions, and offers opportunities for studying friction at the nanoscale from resonant frequency measurements.

**Introduction**

Clamping conditions govern the dynamics of all vibrational systems. This principle can be intuitively understood by listening to string instruments. For example, the distinctive timbre of the sitar, an instrument from India, originates from the modulations of the distance between the clamping points of the strings as the strings vibrate[1]. At the lower boundary of the length scale, nanomechanical resonators are also vibrational systems that are often described as scaled-down versions of string instruments. There, clamping can take multiple forms, ranging from simple, fixed clamping[2-6] to elaborate soft clamping engineered to minimize dissipation in micromachined resonators[7-10]. A nanomechanical resonator sliding on its clamping points embodies a different type of clamping conditions, thus far unexplored at the nanoscale, and reminiscent of those of the sitar. Because it is difficult to realize such a sliding resonator using micromachining, which is better suited for monolithic devices, we consider instead resonators made by transferring a thin membrane of few-layer graphene (FLG) onto a pre-fabricated substrate[11-13]. Such two-dimensional (2-D) resonators have attracted attention for their use as sensors[14], parametric resonators[15] and playgrounds for intermodal vibration engineering[12, 16]. Usually, they are clamped to their support firmly. However, the fact that the membrane is simply deposited on top of its support gives it the possibility to slide on it.

Here, we present measurements of an unconventional, yet robust and controllable dynamics in FLG resonators. This dynamics features vibrational resonant frequencies that draw a loop as a quasi-static pulling force, induced by a gate voltage, is slowly increased and decreased again. Moreover, the frequency loop can be controlled by adjusting the rate at which and the range over which the gate voltage is stepped. We demonstrate that such a dynamics can be explained by a sliding membrane, which breaks with the tradition of simply clamped resonators with fixed boundary conditions and offers additional vibrational degrees of freedom. The sliding occurs in response to the quasi-static force that pulls the membrane into the trench over which it is suspended, instead of simply stretching it as is commonly observed in other nanomechanical resonators. As the membrane is pulled inwards, the length of its suspended part effectively increases, which modifies the resonant frequencies of flexural vibrations. The sliding is slow on the scale of the time needed to measure the frequency response of the vibrational modes, so resonant frequencies can still be estimated from the spectrum of the response. Interestingly, the sliding is reversible –decreasing the pulling force makes the membrane slide outwards. For a given vibrational mode, we find that the area within the loop drawn by the resonant frequency in the space spanned by frequency and gate voltage is a measure of the friction energy dissipated as the membrane slides back and forth on its support. Our work may thus represent a novel approach to quantifying nanoscale friction at cryogenic temperatures.

**Results**

We study the dynamics of our resonators by measuring the resonant frequencies of their vibrational modes. 2-D resonators are known for the large tunability of their resonant frequencies[3, 16-19]. This tunability is ordinarily achieved by suspending the membrane over a gate electrode and subjecting the membrane to a pulling force with a dc voltage $V_G$ applied to the gate. In general, the resonant frequency of a given mode is simply determined by strain within and the dimensions of the suspended membrane. In the presence of nonzero $V_G$, electrostatic pressure directly couples to strain owing to the small bending rigidity of the membrane[20]. With the clamping points fixed, $V_G$ is the sole frequency tuning knob. Our devices have a similar structure (Figs. 1a, b). They are built on a substrate patterned with a source and a drain electrode, and with a gate electrode at the bottom of a trench over which FLG can freely vibrate. FLG is transferred onto this pre-patterned substrate. The part of FLG that is suspended over the gate is the resonator, while the parts in contact with source and drain are meant to clamp FLG to its nonvibrating edges. We obtain the mechanical response of the resonator by measuring an electromechanical current $I$ as a function of the frequency $f_d$ of a driving force (see Methods). All our measurements are carried out at a temperature of $\approx$ 300 mK in a vacuum of $\approx$ $10^{-7}$ Torr. Figure 1c shows the resonant frequencies $f$, at which $I(f_d)$ peaks, as a function of $V_G$ for two vibrational modes hosted by the device pictured in Fig. 1a. Measurements are done by sweeping $f_d$ at fixed $V_G$, then stepping $V_G$ and sweeping $f_d$ again. We observe that the dependences of $f$ on $V_G$ are strongly asymmetric with respect to $V_G = 0$. This behavior is unexpected, because changes in resonant frequency are caused by changes in strain that only depend on $|V_G|$ (Ref. [20]). We have verified that this asymmetry exists whether $f_d$ is swept upwards or downwards, which allows us to rule out bistabilities associated with mechanical nonlinearities[21] as the origin of the asymmetry. We have also verified that the dependence of the conductance of the device on $V_G$ is the same for increasing and decreasing $V_G$ (Supplementary Information, Section S1), which rules out hysteretic behaviors in our measurement readout[22]. We then focus on the upper frequency branch in Fig. 1c, whose asymmetry is more pronounced, and measure it upon increasing and decreasing $V_G$ (Fig. 1d). There, we find the surprising result that the two measurements are mirror images of each other, $f(V_G, \rightarrow) = f(-V_G, \leftarrow)$, where $\rightarrow$ and $\leftarrow$ indicate whether $V_G$ is increased or decreased. This result shows that the direction along which $V_G$ is stepped is key to understanding the dynamics of our resonator[23].

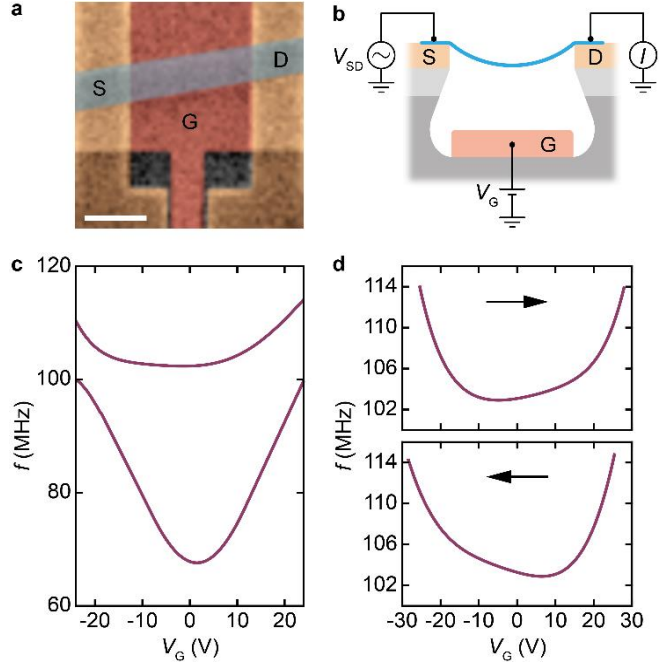

**Figure 1. Unconventional resonant frequency tuning spectra in few-layer graphene (FLG) resonators.** (a) Colorized scanning electron microscope (SEM) image of the device. FLG (blue shaded stripe) is connected to source (S) and drain (D) electrodes and is suspended over a gate electrode (G). Scale bar: 1 micrometer. (b) Schematic of the device. A frequency modulated voltage $V_{SD} = V_0 \cos[2\pi f_d t + (f_\Delta/f_m)\sin 2\pi f_m t]$ is applied between S and D, where $f_d$ is the drive frequency, $f_m = 1.37$ kHz is the modulation frequency, and $f_\Delta/f_m \approx 75$ (see Methods). A dc voltage $V_G$ is applied to G. A drain current $I$ at frequency $f_m$ is measured. (c) Resonant frequency $f$ of the first and second vibrational modes of the resonator shown in (a) as a function of $V_G$. (d) $f$ as a function of increasing (upper panel) and decreasing $V_G$ (lower panel). Arrows indicate the stepping direction of $V_G$. The drive power is -39 dBm in (c) and (d).

To elucidate the relationship between $f$ and the stepping direction of $V_G$, we measure the response of the resonator and its resonant frequency $f$ over narrow $V_G$ ranges in three consecutive stages. Namely, having set $V_G = 20$ V and verified that the response is stable, we sweep $f_d$ and gradually step $V_G$ up to $\approx 30$ V, which results in an increase of $f$ (stage 1 in Fig. 2a). We then immediately reverse the stepping direction. Interestingly, instead of simply following the same path, $f$ first remains constant before decreasing along a shifted path (stage 2 in Fig. 2a). Reversing the stepping direction again at $V_G = 25$ V, $f$ remains constant again and then increases along the same path as in stage 1 (stage 3 in Fig. 2a). The left panel in Fig. 2a shows the superimposed spectra from stages 1-3, revealing a closed frequency loop. To our knowledge, such a phenomenon has not been reported in a nanomechanical resonator thus far. This behavior does not depend on any particular $V_G$ range. We demonstrate this in Fig. 2b, where we cycle $V_G$ between 20 V and 24, 26, 28 and 30 V, and obtain in each case a frequency loop. Overlaying these loops in Fig. 2c makes it clear that $f(V_G)$ follows one path as $V_G$ increases and another, shifted path as $V_G$ decreases. Moreover, reversing the direction of $V_G$ is always followed by a plateau in $f$.

We also find that the shape of frequency loops can be tuned using the rate at which $V_G$ is stepped. Figure

3 shows the response of the resonator as a function of swept $f_d$ and stepped $V_G$ for different rates $dV_G/dt$, where $t$ is time. Within a loop, the largest $V_G$ shift between the two $f(V_G)$ paths defines the width of the loop $\Delta V_G$. As shown in Fig. 3b, $\Delta V_G$ increases with $dV_G/dt$ nonlinearly –faster stepping rates yield wider loops, and $\Delta V_G$ tends to saturate at large $dV_G/dt$. More information about the stepping rate $dV_G/dt$ can be found in Supplementary Information, Sections S2 and S3.

Thus far, frequency loops in nanomechanical resonators have been accounted for by non-mechanical models. For example, applying a dc voltage between source and drain in $MoS_2$ resonators and sweeping the voltage through a cycle produces a resonant frequency loop, which results from changes in strain induced by a phase transition[24]. Applying a magnetic field to $CrI_3$ resonators and sweeping the field up and down also yields a resonant frequency loop due to magnetostriction[25]. In superconducting resonators, applying a magnetic field creates a Lorentz force on vortices that stresses the lattice, producing frequency loops as the field is swept up and down[26]. In these three examples, frequency loops signal an unconventional coupling between field and strain whose origin is a hysteretic subsystem embedded within the resonator. In the absence of a hysteretic process, and excluding any delayed response caused by adsorption-desorption processes[27], measuring resonant frequencies over a certain range of $V_G$ should produce the same result irrespective of whether $V_G$ is increased or decreased. In nanoresonators that are purely mechanical, whose vibrations are not coupled to any hysteretic subsystem, no frequency loop is expected, and thus far, none has been found.

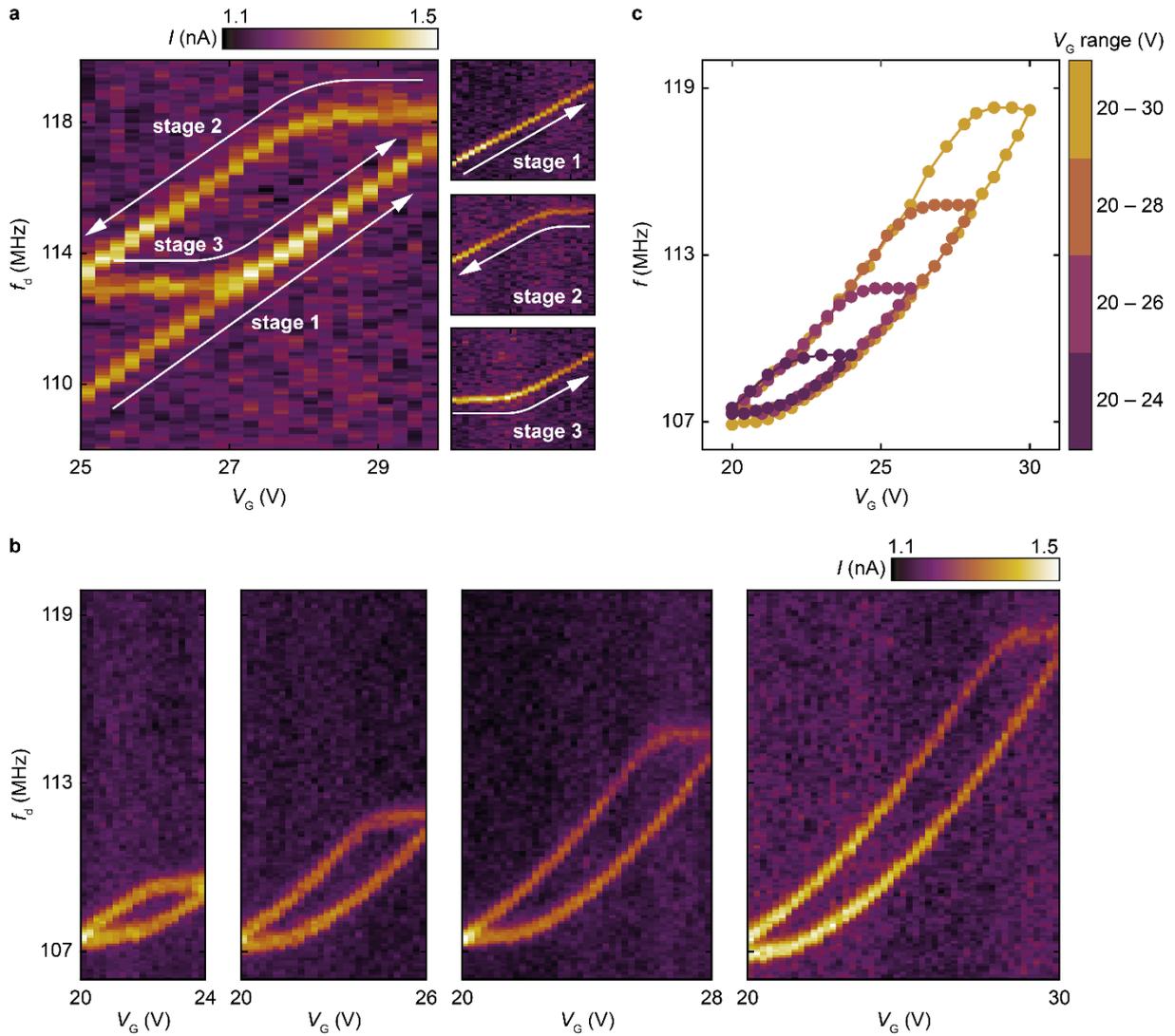

**Figure 2. Resonant frequency loops and their dependence on $V_G$ range.** (a) Frequency response of the mode whose resonant frequency is shown in Fig. 1d as a function of $f_d$ and $V_G$. The response is extracted from the spectrum of current $I$ which peaks at the resonant frequency. In stage 1, $V_G$ is stepped from 20 V up to $\approx$ 30 V. In stage 2, $V_G$ is stepped from $\approx$ 30 V down to 25 V. In stage 3, $V_G$ is stepped from 25 V up to $\approx$ 30 V again. Stages 1-3 are done sequentially and in one go. The left panel shows the superimposed spectra from stages 1-3. (b) Frequency response measured upon increasing and decreasing $V_G$ over various $V_G$ ranges. The panels are made by superimposing data measured with increasing $V_G$ and data measured with decreasing $V_G$. (c) Superimposed dependences of the resonant frequency $f$ on $V_G$ extracted from the four panels in (b).

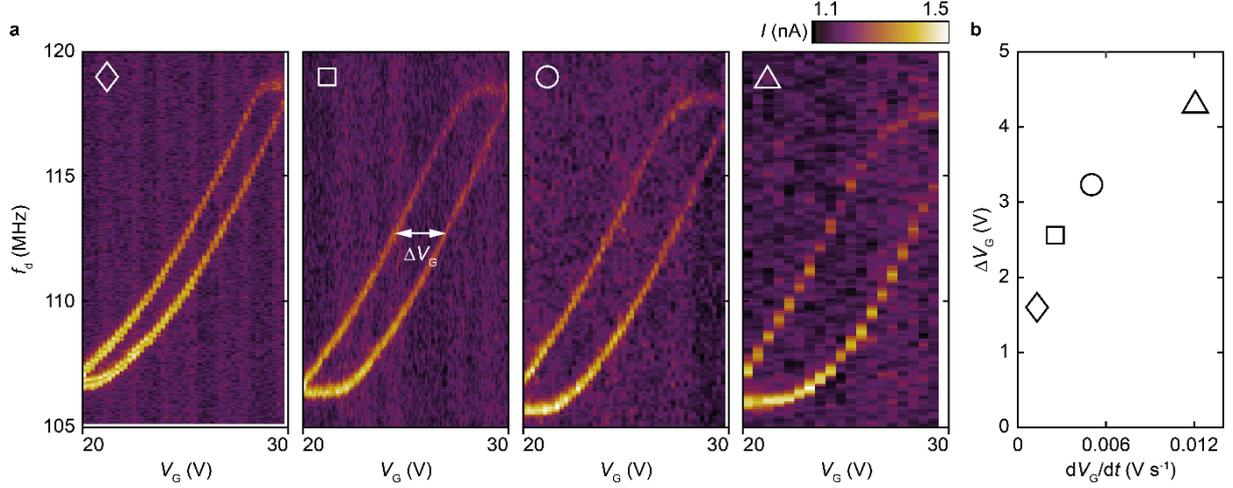

**Figure 3. Resonant frequency loops and their dependence on stepping rate $dV_G/dt$.** (a) Response of the mode shown in Fig. 1d as a function of $f_d$ and $V_G$ for four different stepping rates. (b) Width of frequency loops $\Delta V_G$ as a function of $dV_G/dt$. Marker shape identifies the data set in (a) from which $\Delta V_G$ is extracted.

Understanding the behavior displayed by our resonators, namely their intriguing resonant frequency dependence on $V_G$, calls for an unconventional model. Indeed, we can safely rule out common mechanical phenomena that would also yield frequency loops, such as conservative nonlinearities[21, 28], Euler instabilities[21], and viscoelasticity[29] of graphene. Nonlinearities can be disregarded because the shape of our frequency loops does not depend on the amplitude of the driving force[30] (Supplementary Information, Section S4). Euler instabilities are observed in buckled beams[31], while graphene resonators behave as membranes instead. A viscoelastic graphene membrane, possibly caused by fabrication residues, would not produce a resonant frequency plateau as the stepping direction of $V_G$ is reversed (Supplementary Information, Sections S5 and S6).

A model that may account for our measurements, however, is one that involves unconventional boundary conditions. Namely, we assume that the membrane reversibly slides on the supporting substrate in response to the electrostatic pulling force between the membrane and the gate (Fig. 4a). In turn, this sliding motion modulates the spring constant of the resonator simply by changing the length of the suspended membrane, thereby modifying its strain, hence its resonant frequency. We model the sliding motion with a spring and a dashpot[32] attached to each of the two supporting edges of the membrane (Fig. 4b). Accordingly, the tension $T$ at the edges of the suspended membrane can be described by the Voigt-Kelvin constitutive relationship:

$$T = kq + c\frac{dq}{dt}, \qquad (1)$$

where $k$ and $c$ are the spring constant of the spring and the damping coefficient of the dashpot, respectively. $q$ is the elongation of the spring and dashpot system. The spring extends in response to an increase in the pulling force, feeding extra length of membrane into the suspended area (Fig. 4c). The suspended length increases over a time scale set by the dashpot, which guarantees that $f$ changes in response to a change in $V_G$ with a delay:

$$f = \frac{1}{2\pi}\sqrt{\frac{8ES}{m_{\text{eff}}L}\left(\frac{3}{4}\frac{x^2}{L^2} - 2\frac{q}{L} + \epsilon_0\right)}. \qquad (2)$$

Here, $E$ is Young's modulus, $S$ is the cross-sectional area of the membrane, $\epsilon_0$ is the built-in strain, $L$ is the length of the trench over which graphene is suspended, $x$ is the maximum displacement in the direction perpendicular to the membrane and $m_{\text{eff}}$ is the effective mass of the vibrational mode. $q$ and $x$ can be determined by the following equations (Supplementary Information, Section S7):

$$\frac{dq}{dt} = -\left(\frac{k}{c} + 2\frac{ES}{cL}\right)q + \frac{1}{4}\frac{ES}{cL^2}x^2 + \frac{ES}{c}\epsilon_0, \tag{3}$$

$$\frac{d^2x}{dt^2} = \frac{C'V_G^2(t)}{2m_{\text{eff}}} - \frac{8ES}{m_{\text{eff}}L}\left(\frac{1}{4}\frac{x^2}{L^2} - 2\frac{q}{L} + \epsilon_0\right)x. \tag{4}$$

Here, $C'$ is the first derivative of the gate capacitance with respect to displacement in the vertical direction. We use Eqs. (2-4) to calculate the frequency loops shown in Fig. 4. Using realistic parameters, this phenomenological model reproduces well the shape of frequency loops in the experimental ranges of $V_G$ (the four leftmost panels in Fig. 4d) and the two shifted paths that $f(V_G)$ follows irrespective of these ranges (the rightmost panel in Fig. 4d). Moreover, the model reproduces the width of the frequency loops $\Delta V_G$ as a function of rate $dV_G/dt$ (Fig. 4e). Given that the mass of the resonator and the capacitance to gate can be estimated, three free parameters are needed to reproduce the data. These are the membrane built-in strain $\epsilon_0 = 2.8 \times 10^{-4}$, and the spring constant $k = 8.9 \times 10^2$ kg s$^{-2}$ and the damping coefficient $c = 5.3 \times 10^5$ kg s$^{-1}$ of the spring and dashpot system. Within our model, we find that the frequency loop area $-\oint f\, dV_G$ is proportional to sliding losses $\oint c\frac{dq}{dt}dq$. The ratio of the latter to the former is $\approx 1.01 \times 10^{-24}$ kg m$^2$ s$^{-1}$ V$^{-1}$ and does not depend on the stepping rate $dV_G/dt$ (Supplementary Information, Sections S8 and S9). For one $V_G$ cycle between 20 and 30 V, we estimate sliding losses to be $\approx 4.83 \times 10^{-17}$ J, elongation $q \approx 0.8$ nm and damping force $c\, dq/dt \approx 10^{-7}$ N (Supplementary Information, Sections S7, and S8). From these, we estimate that losses per graphene unit cell amount to $\approx 6.09 \times 10^{-21}$ J, given the width of the membrane of $\approx 500$ nm and assuming that the supported membrane is fully in contact with the substrate. We have observed frequency loops in two devices, referred to as Device A and Device B. We present data from Device A here and show data from Device B in Supplementary Information, Section S10. Although a much larger $c$ is estimated for Device B, the sliding losses per graphene unit cell, $6.72 \times 10^{-21}$ J, are close to those estimated for Device A.

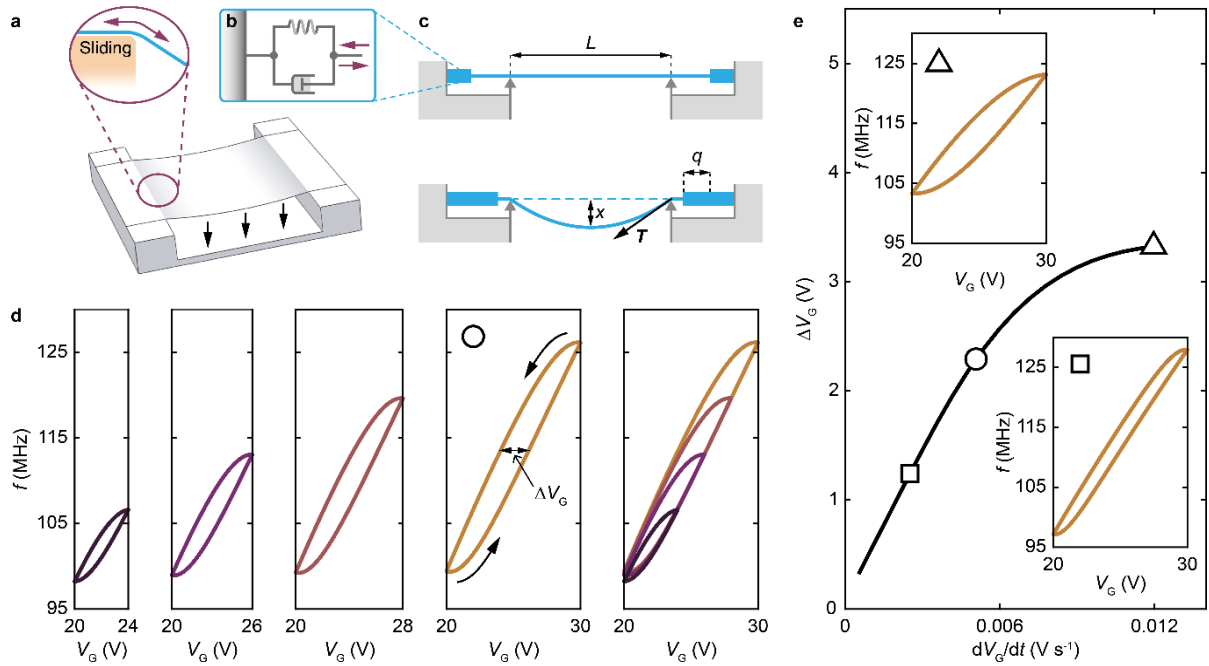

**Figure 4. Mechanical model reproducing frequency loops.** (a) We assume that the supported membrane slides over the substrate controllably (double headed arrows in the inset) as a varying electrostatic pressure (vertical arrows) modulates its vertical displacement. (b) We model this sliding by assuming that the clamping area behaves as a spring and dashpot system. (c) Schematic of the sliding model. Extending spring feeds extra membrane into the suspended area, making the resonator longer and lowering the resonant frequency $f$. Contracting spring pulls the membrane away from the suspended area, shortening the resonator and increasing $f$. (d) Calculated $f$ as a function of $V_G$ for different $V_G$ ranges. Arrows indicate the path followed by $f(V_G)$. The rightmost panel in (d) displays the superimposed frequency loops shown in the four leftmost panels. (e) Calculated width of frequency loop $\Delta V_G$ as a function of $dV_G/dt$. Marker shape identifies the data set ((d) and insets to (e)) from which $\Delta V_G$ is extracted.

## Discussion

The above analysis indicates that our work may contribute to the research efforts on friction between 2-D materials and solid surfaces[33, 34]. First, the friction force we measure depends on the rate $dV_G/dt$, in a way that is reminiscent of the scanning rate dependence of the friction force between graphite and the tip of an atomic force microscope (AFM) probe[35]. Second, our estimate for sliding losses per graphene unit cell is close to $\approx 10^{-20}$ J obtained from those same friction force microscopy experiments[35]. This can be shown by integrating the friction force over tip displacement in Fig. 3A of Ref. [35] and dividing it by the number of graphene unit cells in contact with the tip during the scan. Moreover, our estimate of the damping force divided by the contact area of $5 \times 10^5$ nm² (given by the product of the width of the graphene flake $W \approx$ 500 nm multiplied by the length of the metal contact $L_e \approx 1$ μm) yields a frictional shear stress $\tau \approx 0.20$ MPa ($\tau \approx 0.17$ MPa for Device B). This estimate is smaller than $\tau$ measured in experiments where metal nanoparticles were pushed on the adsorbate-coated surface of a graphite crystal with the tip of an AFM probe[36]. In those experiments, the friction force $F_{fr}$ between a nanoparticle and graphite was found to scale linearly with the contact area $A_c$ of the nanoparticle, resulting in rather large shear stresses $\tau = F_{\text{fr}}/A_c$ ranging from several to hundreds of MPa (Refs. [37,38]). This linear scaling law and the correspondingly large $\tau$ values are understood to originate from interfacial adsorbates, such as hydrocarbons, that hinder the sliding motion of two contacting surfaces[37,38]. In our devices, the smaller estimate for $\tau$ hints at rather clean interfaces between FLG and the electrode. The occurrence of such clean interfaces may be rare, as fabrication residues are otherwise prone to introduce contamination. It may explain why we observe frequency loops in some but not all the devices we have fabricated (Supplementary Information, Section S11). Systematic studies with devices of various dimensions and controlled interface quality may shed light on the friction-area scaling law[38] in these systems. We believe that comparing our estimates made at 300 mK with estimates from Refs. [35,36] obtained at 300 K is meaningful. Indeed, friction forces measured between the tip of an AFM probe and atomically flat $MoS_2$ were found to increase upon lowering temperature from 300 K and reached a plateau near 220 K (Ref. [39]). However, in the case of an artificially roughened $MoS_2$ surface, a much weaker temperature dependence was found below 300 K (Ref. [39]). In both cases, no measurements were made below 100 K. It is not yet known how friction forces should behave in the case of FLG deposited on electrodes at such low temperatures. Overall, our work opens up possibilities for measuring frictional characteristics of 2-D materials at cryogenic temperatures. It also invites future research on friction based on 2-D mechanical resonators held by atomically flat supports, in which case superlubricity[40,41] may confer resonators unusual properties.

Finally, we discuss the reversibility of the sliding motion. Reversibility upon gate tuning is an important feature that is needed in our model to explain our measured frequency loops. It distinguishes our measurements from previously reported frequency instabilities in nanotube resonators[42,43]. Here we discuss a possible origin for the spring that makes sliding reversible. Figure 5a shows an enlarged area surrounding the resonator. It reveals the presence of two neighboring resonators, labelled as $R_L$ and $R_R$, on the left and right sides of it. The three resonators are mechanically connected together because they are made of the same FLG flake deposited over three parallel trenches. We surmise that the dynamics of the resonator in the middle, which is the one we investigate here, is directly influenced by the quasi-static displacement of $R_L$ and $R_R$ according to the simple mechanism that follows. As extra length is fed into the middle trench (Fig. 5b, state (1) to state (2)), strain within $R_L$ and $R_R$ increases. The concomitant in-plane sliding and out-of-plane displacement of $R_L$ and $R_R$ give rise to a tension within the resonators (Fig. 5c). This tension acts as a restoring force (Fig. 5b, state (2) to state (3) and back to state (1)). In Supplementary Information, Section S12, we refine our sliding model by considering the effect of $R_L$ and $R_R$. Using our extended model to account for our measured data, we estimate that $k \approx 9.9 \times 10^2$ kg s$^{-2}$. This is close to $k \approx 8.9 \times 10^2$ kg s$^{-2}$ obtained with the simpler model, which shows that the two models are consistent with one another while the extended one provides insight into the reversible sliding (both models yield similar values for $k$ in Device B as well).

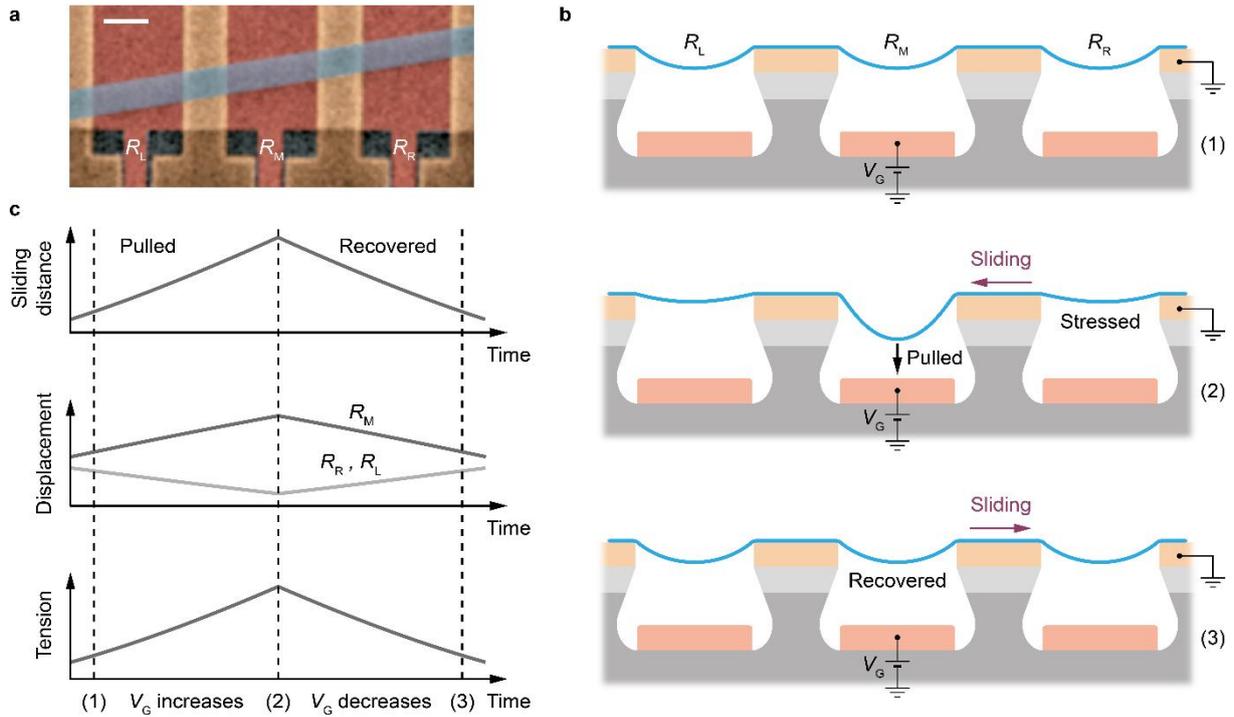

**Figure 5. Physics of the reversible sliding.** (a) SEM image obtained by zooming out of the area shown in Fig. 1a, revealing three mechanically connected resonators, $R_L$, $R_M$, and $R_R$. The resonator investigated in the text is the middle one, labelled as $R_M$. Scale bar: 1 micrometer. (b) Schematic of the reversible sliding. As extra membrane length is fed into the middle trench in response to electrostatic pulling (state (1) to state (2)), strain within $R_L$ and $R_R$ increases. The concomitant in-plane sliding and out-of-plane displacement of $R_L$ and $R_R$ give rise to a tension within the resonators, which acts as a restoring force (state (2) to state (3)). (c) Schematic plot showing the sliding distance, the out-of-plane displacement and the tension within the resonators upon increasing and decreasing $V_G$.

In summary, we demonstrate that the resonant frequency of nanomechanical resonators can be tuned along a loop by cycling a gate voltage. This is a robust effect that is not limited to certain ranges of gate voltages. This is also a subtle effect that we observe in certain devices only. We propose a simple mechanical model to account for it, whereby the resonant frequency of the suspended resonator is modulated by the sliding motion of the membrane on the substrate. We estimate losses incurred as a result of this sliding motion, which are close to measured frictional dissipation between graphite and the tip of an AFM probe. Our work opens up interesting possibilities for studying friction between 2-D materials and their supporting substrates from resonant frequency measurements of their vibrating modes. Our work also offers new perspectives in nanomechanics. Namely, because the sliding part of the membrane acts on the dynamics of the suspended part in an otherwise continuous system, our devices may be the first realization of compliant mechanisms at the nanoscale. Further, while nanomechanical resonators have in common to be firmly anchored to their support, our devices break with tradition and feature sliding clamping areas that enable time-varying boundary conditions. As such, our resonators have much in common with certain musical instruments[1] from Asia, e.g. the sitar, the tanpura, the guqin, and the shamisen, whose distinctive timbres are related to the time-dependent clamping configuration of their strings.

## Methods

### Sample Fabrication

We use a highly resistive silicon wafer coated with a 1000-nm thick silicon oxide layer as a substrate. We first deposit a 50 nm thick layer of $SiN_x$ onto the substrate via low-pressure chemical vapor deposition (LPCVD). Following electron beam lithography (EBL), a trench is defined by a two-step etching process using fluorine-based plasma and hydrofluoric acid, respectively. The total etching depth is approximately 170 nm. After a second EBL step, 3 nm of titanium and 20 nm of gold are evaporated onto the substrate. Using the undercut formed by the $SiN_x$ and $SiO_2$ layers, the evaporated metal can be self-aligned to form three electrodes. Two contacts serve as source and drain for electrical contacts. An electrode in the trench serves as a gate for electrical tuning. Finally, a few-layer graphene ribbon, exfoliated on a polydimethylsiloxane (PDMS) stamp, is transferred onto the trench[44, 45]. The suspended part of the device investigated in the main text (Device A) has a length of 1.82 μm and a width of 0.52 μm (measured from the SEM image). The additional device (Device B) has a length of 1.98 μm and a width of 3.24 μm (Supplementary Information, Section S10).

### Measurement setup

To detect the mechanical resonance of the nanomechanical resonator, a frequency modulation (FM) mixing technique[46, 47] is employed to actuate and detect the mechanical vibrations. The FM signal has the form $V_{SD}(t) = V_0 \cos[2\pi f_d t + (f_\Delta/f_m) \sin 2\pi f_m t]$, where $V_0$ is the amplitude of the drive voltage, $f_d$ is the drive frequency, $f_\Delta$ is the deviation frequency (typically 103 kHz), and $f_m$ is the modulation frequency (typically 1.37 kHz). This technique provides both a capacitive force at $f_d$ that drives vibrations and a drain current $I$ at frequency $f_m$. We use a lock-in amplifier to detect $I$ at the drain electrode. The transduced mixing current $I$ is proportional to $|\partial \text{Re}[z]/\partial f_d|$, with $\text{Re}[z]$ the real part of the vibrational amplitude, thus allowing us to investigate the mechanical vibrations of the resonator.

### Author contributions

Y. Y. and Z.-Z. Z. contributed equally to this work. X.-X. S. and G.-P. G. designed the research. Z.-Z. Z. and Z.-J. S. fabricated the device. Y. Y., Z.-Z. Z. and X.-X. S. performed low-temperature measurements. Y. Y., J. M. and X.-X. S. provided the theoretical model. Y. Y., Z.-Z. Z., J. M. and X.-X. S. analyzed the data. Y. Y., Z.-Z. Z., J. M. and X.-X. S. co-wrote the manuscript with inputs from all other authors. All the authors contributed to discussions.

### Competing interests

The authors declare no competing interests.

### Data availability

The data that support the findings of this study are available in Zenodo (https://doi.org/10.5281/zenodo.7049075).

**Acknowledgements**

The authors would like to thank Feng-Chao Wang, Antoine Reserbat-Plantey and Vincent Bouchiat for fruitful discussions. This work was supported by the National Key Research and Development Program of China (Grant No. 2016YFA0301700 (G.-P. G.)), the National Natural Science Foundation of China (Grant Nos. 61904171 (Z.-Z. Z.), 11904351 (X.-X. S.), 11625419 (G.-P. G.), 12034018 (G.-P. G.), 62074107 (J. M.), and 62150710547 (J. M.)), the Anhui Initiative in Quantum Information Technologies (Grant No. AHY080000 (G.-P. G.)), the Anhui Provincial Natural Science Foundation (Grant No. 2008085QF310 (Z.-Z. Z.)), and the International Cooperation and Exchange of the National Natural Science Foundation of China NSFC-STINT (Grant No. 61811530020 (J. M.)). This work was partially carried out at the USTC Center for Micro and Nanoscale Research and Fabrication.